%% file: Sample_Size_arxiv.tex
\documentclass[11pt]{article}
\usepackage[utf8]{inputenc}
\usepackage[margin=1in]{geometry}
\usepackage{amsmath}
\usepackage{physics}
\usepackage{xcolor}
\usepackage{dirtytalk}
\usepackage{graphicx}
\usepackage{listings}
\usepackage{lscape}
\usepackage{cite}
\usepackage{booktabs}

\usepackage[colorlinks=true,linkcolor=purple,citecolor=blue]{hyperref}
\let\originaleqref\eqref
\renewcommand{\eqref}{equation~\originaleqref}

\DeclareMathOperator*{\argminA}{arg\,min} 

\providecommand{\keywords}[1]
{
  \small	
  \textbf{\textit{Keywords---}} #1
}

\title{Minimum Sample Size Calculation for Multivariable Regression of Continuous Outcomes in Chemometrics for Astrobiology and Planetary Science}
\author{Menelaos Konstantinidis$^{1,2}$,Emmanuel A. Lalla$^{3,4,*}$, Sofia Julve Gonzalez$^{5}$, J. Manrique$^{5}$\\
Guillermo Lopez-Reyes$^{5}$, Andrew Barlow$^{6}$,\\Elissa Sawyers$^{3}$, Belen Barrios$^{3}$,\\ Michael G. Daly$^{3}$, \\
$^1$ Institute of Health Policy, Management and Evaluation, University of Toronto, ON, Canada \\
$^2$ Li Ka Shing Knowledge Institute, St. Michael's Hospital, Unity Health Toronto, ON, Canada \\
$^3$Centre for Research in Earth and Space Science, York University, Toronto, ON, Canada \\
$^4$Teledyne Geospatial, Vaughan, ON, Canada \\
$^5$University of Valladolid, Valladolid, Spain\\
$^6$Institute for Aerospace Studies, University of Toronto, Toronto, ON, Canada\\
$^*$Corresponding author email: elalla@yorku.ca\\
}

\date{}

\begin{document}

\maketitle

\begin{abstract}
Over the last few decades, prediction models have become a fundamental tool in statistics, chemometrics, and related fields. However, to ensure that such models have high value, the inferences that they generate must be reliable. In this regard, the internal validity of a prediction model might be threatened if it is not calibrated with a sufficiently large sample size, as problems such as overfitting may occur. Such situations would be highly problematic in many fields, including space science, as the resulting inferences from prediction models often inform scientific inquiry about planetary bodies such as Mars. Therefore, to better inform the development of prediction models, we applied a theory-based guidance from biomedical domain for establishing what the minimum sample size is under a range of conditions for continuous outcomes. This study aims to disseminate existing research criteria in biomedical research to a broader audience, specifically focusing on their potential applicability and utility within the field of chemometrics. As such, the paper emphasizes the importance of interdisciplinarity, bridging the gap between the medical domain and chemometrics. Lastly, we provide several examples of work in the context of space science. This work will be the foundation for more evidence-based model development and ensure rigorous predictive modelling in the search for life and possible habitable environments. 
\end{abstract}


\section{Introduction}

The burgeoning field of astrobiology seeks to understand the origins, evolution, distribution, and future of life in the universe. Within this multidisciplinary science, chemometrics plays a pivotal role in deciphering the complex chemical signatures that might indicate the presence of past or present biological activity.

The detection and analysis of biosignatures – life indicators – relies heavily on the sophisticated analytical techniques developed within chemometrics. For instance, the calibration models employed in the analysis of Laser-induced Breakdown Spectroscopy (LIBS) data from Mars exploration rovers are equally applicable to identifying organic compounds and minerals that could be indicative of biological processes \cite{Cremers2013, LMM,Konstantinidis2019}.

Throughout the last several decades, prediction models answering increasingly complex questions have proliferated in the field of chemometrics and related disciplines such as statistics, biostatistics, and machine learning. At its core, a prediction model estimates the expected value of a response variable conditioned upon one or more explanatory variables. In the simplest cases, this may be ordinary least square regression for continuous response variables and logistic regression for categorical response variables. Subsequently, the result of a prediction model is an equation that can be used to provide a predicted expected value for some new observation.

As an example, in the context of planetary exploration and search of life, atomic emissions from Laser-induced Breakdown spectra (LIBS), over a series of spectral channels, may be used to predict the abundance of elements in a geological sample \cite{Cremers2013, LMM,Konstantinidis2019}. Regression models using independent component analysis or partial least square regression are often calibrated over a set of samples, from which the abundance of key elements is predicted, thereby allowing the estimation of elemental abundance from unknown samples. This has become a particularly appealing approach in the context of Mars with instruments such as ChemCam and SuperCam, for which semi-autonomous processes (including elemental predictions) have been implemented \cite{Clegg2009,Clegg2017,Wiens2020}.

In some fields, the proliferation of prediction models has been made possible by the increasing ease with which data is generated; however, in other fields (e.g., planetary exploration), generating data remains a laborious and expensive process, especially at the calibration stage. Consequently, in cases where the latter holds, systematic development of a calibration set may be forgone, and instead, a smaller convenience set may be used. This may in turn exacerbate or give rise to methodological problems such as overfitting, optimism, and insufficient precision of parameter estimates. 

Overfitting refers to fitting a statistical model in which there are too many degrees of freedom \cite{Steyerberg2009}. This may in turn lead to false assurance of the fitted model’s performance on new observations. In contrast, optimism refers to the difference between true performance and observed performance \cite{Steyerberg2009,Riley2018}. In other words, it is the difference in the performance between a model’s performance in a calibration set, and how it would perform on the population. It is therefore necessary for the sample size of a model to be large enough so that underlying problems such as overfitting and optimism are mitigated when estimating key estimands (e.g., RMSE). Precisely how large, is a question that we will endeavour to answer in this article. For example, LIBS spectra from the ChemCam instrument have 6144 spectral channels, potentially risking overfitting if the sample size is insufficient. 

Despite the clear need for systematic sample size calculations in the development of a dataset, in the chemometric literature, this is often considered only heuristically on a post-hoc basis, that is, based on the observed performance of a model with respect to some criterion. In other cases, rules of thumb are often used, such as the events per variable (i.e., 10 events per predictor in logistic regression); although such guidance has been shown to be unsubstantiated \cite{Smeden2016,Smeden2018}. The biomedical literature has evaluated the minimum sample size problem at length, and it is largely from this body of evidence that we draw from to guide chemometricians. 

In this article we will focus on continuous outcome multivariable regression models. Largely based on the biostatistical literature \cite{Steyerberg2009,Riley2018,Smeden2018,Collins2015,Moons2015,Steyerberg2000,Harrell1996,Steyerberg2001a}, we will provide three criteria by which to estimate the minimum sample size required for a prediction such that the possibility of overfitting is minimized and precision of key estimands may obtained. Moreover, we intend to extent these criteria methods from \cite{Riley2018}, which never have been considered Planetary science. The rest of the article will proceed as follows. In section 2, we will present the mathematical framework for the sample size calculation of prediction models using the three proposed criteria. In section 3, an extension of these calculations to estimate the root mean squared error (RMSE). In section 4, we will provide worked example from LIBS in the context of Mars. Lastly, in section 5, we will provide a discussion of presented sample size calculations and how such calculations should be reported in prediction models. Moreover, we are providing an appendix clear definitions and explanations
of key statistical concepts and parameters introduced in this research
approach for astrobiology.

\section{Sample Size Calculations}
In this article, we consider the minimum sample size required for the calibration of a predictive model based on three criteria, each designed to either minimize the risk of overfitting and optimism, or maximize the precision in the estimation of one or more parameters. Moreover, we followed a similar notation as established in \cite{Steyerberg2009,Riley2018,Smeden2018,Collins2015,Moons2015,Steyerberg2000,Harrell1996,Steyerberg2001a}. The first two criteria we present (the difference between $R^2$ and $R_{adj}^2$, and the global shrinkage factor) provide an absolute and relative measure against which to minimize overfitting while accounting for optimism. The third criterion – precision of the residual standard deviation – provides us with relative measure against which to estimate the residual standard deviation, which may in turn be used to calculate other estimands (e.g., the confidence interval of regression coefficients or the prediction interval of an outcome).  

Throughout, we consider a continuous regression of the form

\begin{equation}\label{eq:eq.1}
    Y_i=\beta_0+\beta_{1}X_{1}+ \beta_{2}X_{2}+ \dots +e_{i}
\end{equation}
for $i=1, \dots, n$ observations such that $e_{i}\sim N(0,\sigma^{2})$

\subsection{$R^{2}$ and $R_{adj}^{2}$}
The first criterion we suggest is an absolute measure of overfitting, namely the absolute difference between $R^2$ and the adjusted $R^2$, the latter denoted $R_{adj}^2$. Past work has suggested that a minimal sample size to ensure a small degree of overfitting should be calculated such that the difference is small (e.g., at most 0.05) \cite{Riley2018}. Specifically, we define

\begin{equation}\label{eq:eq.2}
    \Delta= R^{2}-R_{adj}^{2}
\end{equation}

We observe that 

\begin{equation}\label{eq:eq.3}
    R_{adj}^{2}=\dfrac{(1-R^{2})(n-1)}{(n-p-1)}
\end{equation}

is an unbiased, optimism-adjusted estimate of $R_{adj}^2$ \cite{Riley2018,Copas1983,Goldberger1964}, where $p$ is the number of predictor parameters. Rearranging, we get an unbiased, optimism-adjusted estimate of $R^2$

\begin{equation}\label{eq:eq.4}
    R^{2}=\dfrac{R_{adj}^{2}(n-p-1)+p}{n-1}
\end{equation}

Subsequently, plugging \hyperref[eq:eq.4]{eq. 4} in \hyperref[eq:eq.2]{eq. 2}, we find that 

\[
    \Delta=\dfrac{p\left(1-R_{adj}^{2}\right)}{n-1}
\]

Rearranging, we get that the minimum sample size is given by 

\begin{equation}\label{eq:eq.5}
    n_{min}=1+\dfrac{p\left(1-R_{adj}^{2}\right)}{\Delta}
\end{equation}

where $\Delta$ is small.

\subsection{The Global Shrinkage Factor}
To account for overfitting in \hyperref[eq:eq.1]{eq. 1}, we may apply a correction known as a Global Shrinkage Factor (GSF). This is an approach whereby, after initial parameter estimation of \hyperref[eq:eq.1]{eq. 1} (e.g., through maximum likelihood estimation), a shrinkage factor $S$ is applied to the estimated predictor coefficients such that predictions made on new observations are done so through the following equation

\begin{equation}\label{eq:eq.6}
    Y_i=\beta_{0}^{'}+S\left(\beta_{1}X_{1}+ \beta_{2}X_{2}+ \dots\right)
\end{equation}
Here, $\beta_{0}^{'}$ is a revised intercept, and $S$ needs to be estimated. The effect of using $S$ is that the parameters are shrunk toward the mean. While there are several ways of estimating $S$, we will proceed with the analytical framework described by (Riley et al., 2019), namely the Copas estimate \cite{Smeden2018,Copas1983,Copas1997}. Specifically, the Copas shrinkage factor $S_{c}$ is defined as 

\begin{equation}\label{eq:eq.7}
    S_{c}=1-\dfrac{p-2}{LR}
\end{equation}

where
\begin{equation}\label{eq:eq.8}
    LR=-2(l_{null}-l_{model})
\end{equation}

such that  $l_{null}$ is the log-likelihood of the model with intercept only (i.e., no predictors), and $l_{model}$ is the full model (i.e., the model with the full set of included predictors). Note that in the estimation of $S_c$, while the full model (i.e., $l_{model}$) may not include all possible predictors, $p$ is explicitly defined as the cardinality of the parameter space (i.e., all possible predictors). It is also worth noting that in defining $p$, we make a distinction between the number of variables and the number of predictors. For example, categorical variables with three categories will have two predictor parameters. In this regard, $p$ is the number of predictor parameters and not the number of variables. 

Thus far we have defined the estimate of the shrinkage factor $S_c$, but recall that our objective is to provide a minimum sample size, with $S_c$ as the criteria. To do so, we first note that 

\begin{equation}\label{eq:eq.9}
    R^{2}=1-\dfrac{\hat{\sigma}^{model}}{\hat{\sigma}^{null}}
\end{equation}

Furthermore, as described in previous work \cite{Steyerberg2001},

\begin{equation}\label{eq:eq.10}
    LR=-n\log(1-R^{2})
\end{equation}

Thus, from \hyperref[eq:eq.8]{eq. 8}, 

\begin{equation}\label{eq:eq.11}
S_{c}=1+\dfrac{p-2}{n\log(1-R^{2})}    
\end{equation}

and applying \hyperref[eq:eq.4]{eq. 4}, 

\begin{equation}\label{eq:eq.12}
    S_{c}=1+\frac{p-2}{n\log\left(\dfrac{\left(n-p-1\right)\left(1-R_{adj}^{2}\right)}{n-1}\right)}
\end{equation}

It then follows that we can find the minimum sample size numerically against some threshold of $S_c$; (Riley et al., 2019) suggest that $S_c$ should be at least 0.9 to minimize overfitting \cite{Riley2018}. Thus, our objective function is as follows:

\begin{equation}\label{eq:eq.13}
    n_{min}=\argminA_n \left\{1+\frac{p-2}{n\log\left(\dfrac{\left(n-p-1\right)\left(1-R_{adj}^{2}\right)}{n-1}\right)}\geq 0.9\right\}
\end{equation}

from which we observe that as $n$ increases, $R_{adj}^{2}$ increases, $n/p$ increases, and $p$ decreases.

\subsection{Residual Standard Deviation}
As described in previous work, the residual standard deviation of a regression is an important estimand, as it allows for standard errors for the intercept and predictor parameters, and by extension the mean predicted outcome and prediction interval at fitted values of X to be estimated \cite{Montgomery2021,Harrell2015,Fox2015,Gelman2020}. Thus, following the work of (Riley et al., 2019), we proceed to define the minimum sample size by which the residual standard deviation can be estimated within a specific margin of error under a 95\% confidence interval for the margin of error in a similar way as \cite{Riley2018}. 

Specifically, we define the multiplicative margin of error (MMOE) by 

\begin{equation}\label{eq:eq.14}
    MMOE=\sqrt{max\left(\dfrac{\chi^{2}_{1-\frac{\alpha}{2},n-p-1}}{n-p-1},\dfrac{n-p-1}{\chi^{2}_{\frac{\alpha}{2},n-p-1}}\right)}
\end{equation}

where $\chi^{2}_{\frac{\alpha}{2},n-p-1}$ is the critical value of the Chi-squared distribution with probability $\alpha/2$ and $(n-p-1)$ degrees of freedom. Thus, for the residual standard error of the full model $\hat{\sigma}_{model}^{2}$, to have a margin of error of no more than say 10\% with a $100(1-\alpha)\%$ confidence interval, we need the MMOE to be at most 1.2. The $n_{min}$ for which this is true at a given value of $p$ and $\alpha$ can be easily solved numerically. The objective function is given by 

\begin{equation}\label{eq:eq.15}
    n_{min}=\argminA_n\left\{\sqrt{max\left(\dfrac{\chi^{2}_{1-\frac{\alpha}{2},n-p-1}}{n-p-1},\dfrac{n-p-1}{\chi^{2}_{\frac{\alpha}{2},n-p-1}}\right)}\leq 1.1\right\}
\end{equation}

\section{Extensions}
Up to now we have delineated three key criteria to find the minimum sample size with which the risk of overfitting is minimized and the residual standard deviation is estimated with a desired degree of precision. However, prior to providing an empirical example, we will provide an extension of these concepts to two additional estimands, namely the root-mean-squared-error (RMSE), and the prediction interval for the mean outcome. 

\subsection{RMSE}
The RMSE is a common way to assess model adequacy based on how precisely a given outcome (over a set of observations) can be estimated. In developing a prediction model, it is therefore important to quantify the necessary sample size such that a given RMSE can be obtained. The RMSE is given by 

\begin{equation}\label{eq:eq.16}
    RMSE(y)=\sum_{i=1}^{n}\sqrt{\dfrac{\left(\hat{y}_{i}-y_{i}\right)^{2}}{n}}
\end{equation}

where $y_{i}$ is the true value of the response for the $i$th observations, and $\hat{y}_{i}$ is the predicted value of the response for the $i$th observation. The RMSE can also be written as 

\begin{equation}\label{eq:eq.17}
    RMSE(y)=\sqrt{\dfrac{\hat{\sigma}_{model}^{2}}{(n-p-1)}}
\end{equation}

or equivalently as

\begin{equation}\label{eq:eq.18}
    RMSE(y)=\sqrt{\dfrac{\hat{\sigma}_{null}^{2}\left(1-R^{2}\right)}{(n-p-1)}}
\end{equation}

A natural extension would be to find a minimum sample size such that the RMSE can be estimated with minimal risk of over-fitting. We observe from \hyperref[eq:eq.18]{eq. 18} that the RMSE is dependent on $R^2$ and $\hat{\sigma}_{null}^{2}$. Since both are post-estimation values, we need some alternative functional form. This is easily resolved in the case of $R^2$, since we can make use of \hyperref[eq:eq.2]{eq. 2} overfitting-adjusted estimate. The case of $\hat{\sigma}_{null}^{2}$ follows in a similar fashion, in that a value must be assumed. Thus, we have the following expression for the overfitting-adjusted RMSE

\begin{equation}\label{eq:eq.19}
    RMSE(y)=\sqrt{\dfrac{\hat{\sigma}_{null}^{2}\left(1-\left(\Delta+R_{adj}^{2}\right)\right)}{n-p-1}}
\end{equation}

Therefore

\begin{equation}\label{eq:eq.20}
    n_{min}=1+p+\dfrac{\hat{\sigma}_{null}^{2}\left(1-\Delta-R_{adj}^{2}\right)}{(RMSE_D)^{2}}
\end{equation}

where $RMSE_D$ is the proposed RMSE.

\subsection{Mean predicted outcome}
Let us assume that we have a regression model with a set of predictors centred at their mean values. It then follows that the fitted intercept $\hat{\beta}_{0}$ corresponds to the outcome, conditioned upon the mean-valued variables. Subsequently, the variance of $\hat{\beta}_{0}$ is given by 

\begin{equation}\label{eq:eq.21}
    var(\hat{\beta}_{0})=\frac{\hat{\sigma}_{model}^{2}}{n}=\frac{\sigma^{2}_{null}\left(1-R^{2}_{adj}\right)}{n}
\end{equation}

subsequently, the 95\% confidence interval for the intercept is 

\begin{equation}\label{eq:eq.22}
\hat{\beta}_{0}\pm \left(t_{1-\frac{\alpha}{2},n-p-1}\sqrt{\frac{\sigma^{2}_{null}\left(1-R^{2}_{adj}\right)}{n}}\right)    
\end{equation}

Thus, for a given value of $R^{2}_{adj}$ and $\sigma^{2}_{null}$, we can solve for the $n_{min}$ that would provide the desired precision of $\hat{\beta}_{0}$.

\section{A Worked Example }
In the context of spectroscopy, the entire spectrum is often not used, due to any number of reasons including computational time, multicollinearity, and matrix effects. Therefore, for the purpose of our illustration, we may assume a reduced space (possibly latent as in the case of Principal Component Regression – see below for more details). Such may be the case in LIBS spectra, where the emissions (or latent space) are correlates of elemental abundance. In this regard, we may reasonably assume that the predictors are mechanistic in their effect on $Y$, and as such, we assume $R_{adj}^{2}=0.5$ (we discuss strategies for the selection of $R_{adj}^{2}$ below). For illustration, we assume that $p=25$. The rest of the section will be dedicated to illustrating the minimum sample size calculations for each of the three criteria, in addition to the RMSE and mean predicted outcome. 

\subsection{$R^{2}$ and $R_{adj}^{2}$}
From \hyperref[eq:eq.5]{eq. 5}, assuming $\Delta=0.05$, we have

\[
n_{min}\geq 1+\frac{25(1-0.5)}{0.05}=251
\]

\subsection{Global Shrinkage Factor}
From \hyperref[eq:eq.13]{eq. 13}

\[
    n_{min}=\argminA_n \left\{1+\frac{23}{n\log\left(\dfrac{\left(n-26\right)\left(1-0.5\right)}{n-1}\right)}\geq 0.9\right\}
\]
from which we find that $n_{min}=269$

\subsection{Residual standard errors}
To find a margin of error on the residual standard deviation of 10\% with a 95\% confidence interval we apply \hyperref[eq:eq.15]{eq. 15} such that

\[
    n_{min}=\argminA_n\left\{\sqrt{max\left(\dfrac{\chi^{2}_{0.975,n-26}}{n-26},\dfrac{n-26}{\chi^{2}_{0.025,n-26}}\right)}\leq 1.1\right\}
\]
from which we find that $n_{min}=233$

\subsection{RMSE}
For the RMSE, we apply \hyperref[eq:eq.20]{eq. 20} such that

\[
n_{min}=1+p+\dfrac{\hat{\sigma}_{null}^{2}\left(1-\Delta-R_{adj}^{2}\right)}{(RMSE_D)^{2}}
\]

From previous work in LIBS for planetary exploration, we used the estimated standard deviation from the LIBS Raman Sensor data for SiO$_2$, specifically, $\sigma_{null}=29.5$ \cite{Konstantinidis2019,LMM}. For illustration, let us suppose that we would like an RMSE of 2\%, that is $RMSE_D=2$. Then,

\[
n_{min}=26+\dfrac{(29.5)^{2}\left(1-0.05-0.5\right)}{(2)^{2}}=124
\]

\subsection{Mean predicted outcome}
To obtain a 95\% confidence interval for the mean outcome – $\hat{\beta}_{0}$, we apply \hyperref[eq:eq.20]{eq. 22}. Moreover, we must assume a mean outcome. To do so, we refer to the LiRS database, for which the mean value of SiO$_2$ is 17.37 \cite{Konstantinidis2019}. Thus, the 95\% confidence interval is given by

\[
17.37\pm \left(t_{0.975,n-26}\sqrt{\frac{(29.5)^{2}\left(1-0.5\right)}{n}}\right)
\]

Suppose that we would like a margin of error of no more than 10\%, then 

we need to find $n$ such that

\[
\left(t_{0.975,n-26}\sqrt{\frac{(29.5)^{2}\left(1-0.5\right)}{n}}\right)\leq1.737
\]
The minimum sample size for which this holds is $n_{min}=557$

\subsection{Summary}
We have applied five criteria for finding the minimum sample size (see Table \ref{table: 1}). Given that, in our example, we wish to find a sample size such that each of the criteria are satisfied, we need to develop a calibration dataset with a largest of the calculated sample sizes, that is, 557 samples. 

\begin{table}[h]
\begin{tabular}{l|llll|l}
\cline{1-6}
     Criterion & p  & $R_{adj}^{2}$ & $\Delta$ & $\sigma_{null}$ & $n_{min}$ \\ \hline
 $R^{2}$ vs. $R_{adj}^{2}$    & 25 & 0.5    & 0.05  & \_    & 251    \\
 Global Shrinkage Factor  & 25 & 0.5    & \_    & \_    & 269    \\
MMOE & 25 & \_     & \_    & \_    & 233    \\
RMSE & 25 & 0.5    & 0.05  & 29.5  & 124    \\
Mean Predicted Outcome  & 25 & 0.5    & \_    & 29.5  & 557    \\ \cline{1-6} 
\end{tabular}
\caption{Summary of parameters and estimated minimum sample sizes for based on the five presented criteria. $\textrm{--}$ indicates no relevant value of a parameter for a given criterion}
\label{table: 1}
\end{table}

\section{Discussion}
\subsection{Overview}
Sample size calculations, while not common in the physical sciences and chemometrics, are a cornerstone of rigorous science in other fields, and we suggest that sample size calculations are a fundamental component for the success and rigorous of prediction models (including those developed in chemometrics). In this article, we presented three criteria (and two extensions) to calculate the minimum sample size to minimize overfitting, model optimism, and allow for precise estimation of model parameters. These are not all new calculations in themselves; however, given the lack of rigorous sample size calculation in the development of prediction models in chemometrics, these criteria are an important contribution to the field. 

Criterion 1 and 2, the difference between $R^2$ and $R_{adj}^{2}$ and the global shrinkage factor, respectively, are particularly appealing approaches as they provide sample size calculations adjusted for optimism and overfitting. Criterion 1 allows the chemometrician to estimate $n_{min}$ corresponding to a small $\Delta$ for a given $p$. We observe, of course, that how small $\Delta$ should be is not well defined. However, drawing from past work, 0.05 has been suggested as a reasonable upper bound \cite{Riley2018}. Furthermore, $R_{adj}^{2}$ must be pre-specified, which can be a challenge. For this we suggest some strategies. 

Firstly, it is suggested that $R_{adj}^{2}$ may be estimated from previous studies or reviews of relevant prediction models. However, it is often the case that such literature does not exist. In such cases, we recommend a conservative approach. Specifically, (Riley et al., 2019) suggested that if the signal-to-noise ratio for any given predictive variable can be assumed to be low, $R_{adj}^{2}$ values ranging from 0.1 to 0.2 may be reasonable \cite{Riley2018}. In such a case, to provide a conservative approach, a $R_{adj}^{2}$ of 0.15 has been suggested when no information is present. However, when the predictors include a mechanistic variable (i.e., one that directly affects the outcome), a less conservative value of $R_{adj}^{2}$ around 0.5 may be taken. This is, of course, context dependent; however, when dealing with spectroscopy data as is often the case in chemometrics and space science (e.g., LIBS or Raman spectroscopy), the spectral channels correspond to emissions or vibrations which are themselves mechanistic in their effect on analyte concentrations (e.g., elemental abundance). Thus, if such may be the case for the chemometrician, a $R_{adj}^{2}$ of 0.5 may be reasonably assumed. 

As with the selection of $\Delta$, similar caveats exist in the selection of the global shrinkage factor; however, it has been suggested that a value above 0.9 is desired  \cite{Riley2018,Harrell2015,Harrell1996}, thereby providing what we may take as a reasonable lower bound.  Similarly, the discussion around the a priori selection of $R_{adj}^{2}$ follows for all of the sample size calculations provided in the previous section. On the other hand, the a priori selection of $\hat{\sigma}_{null}^{2}$ is unlikely to be based on any rules of thumbs as this will be particular to the population at hand. Thus, we suggest that chemometricians refer to past literature, or appeal to domain expertise around the likely variability of the outcome under consideration. 

We now turn briefly to the two extensions, namely, sample size calculations for the RMSE and mean predicted outcome. In both cases, the above discussion applies entirely. It is, however, worth adding that the desired RMSE will vary from one outcome to another (e.g., the desired RMSE in Martian samples for SiO$_2$ may be very different from that of trace elements). Similarly, the assumed mean value for a outcome of interest will vary based on the desired application (e.g., Mars vs. the Moon). 

Furthermore, we would like to remark on a key methodological caveat. The setup provided in \hyperref[eq:eq.1]{eq. 1} is such that the sample size calculations assume a linear model with gaussian residuals. While the sample size calculations that follow from this assumptions are an important contribution, one might wonder how such calculations may generalize to other more frequently used models such as Principal Component Regression (PCR) and Partial Least Squares Regression (PLSR). To such a concern we observe the following. 

In the case of PCR, instead of applying the sample size calculation based on the predictors (from the included explanatory variables), we may do so with respect to the principal components. Thus, the regular rules of linear regression apply here. The difference to be noted is that the residuals in linear regression are often assumed to follow a normal distribution. However, given that we are likely not interested in conducting inference on the principal components but rather on prediction, the normality assumption need not be satisfied. On the other hand, sample size calculations can be applied to PLS by conducting PLS on the shrunken (by Global Shrinkage Factor) predictor space rather than the original predictor space; the case of PLS is a focus of ongoing work. 

Finally, it is important to shed some light on the interplay between minimum sample size calculations and cross-validation. Cross-validation is often erroneously regarded as a sufficient means by which to minimize the risk of overfitting. This is, however, not true in general, as the resulting parameters may still be biased, especially with small sample sizes \cite{afendras2019optimality}. Therefore, it is necessary that for model calibration, a minimum sample size be obtained, after which cross-validation or other approaches may be used for hyperparameter optimization and other objective functions.  

\subsection{Guidance on reporting minimum sample size}
In an endeavour to reduce biases, increase reproducibility, and ultimately ensure more rigorous research, the biomedical community routinely provides sample size calculations prior to the execution of a prospective study (i.e., before data is collected). For example, the minimum sample size may be calculated, so that the prospective data will have sufficient statistical power to detect a statistically significant difference between two groups should such a difference exist. This is typically a requirement for acquisition of funding and publication of prospective studies, and we believe that such standards should be incorporated to increase the rigorour of chemometric studies.
 
A complete proposition on reporting guidelines for studies in chemometrics necessarily exceeds the scope of the present work; however, we would like to propose a few guiding principles for sample size calculations in chemometrics studies. 

\subsubsection{Criteria}

\begin{enumerate}
\item Identify whether this is a model development or a validation study of an existing model
    	
\item Define the estimands that will be reported on (e.g., RMSE, LoD, prediction estimates)
	
\item Identify the relevant sample size criteria for each of the estimand.
	
\item For each criterion in (3), identify parameters that must be defined a priori
	
\item If literature on relevant prediction models exist (e.g., reviews or other prediction models), inform parameter values in (4) by the literature
	
\item If literature does not exist, provide justification around an educated estimate for parameter values (e.g., $R_{adj}^{2}$ of 0.5 for prediction models with mechanistic predictors) 

\item Provide the minimum sample size calculation for each estimand and any additional assumptions that were made
	
\item Provide fitted estimates of estimands and criteria to establish whether the sample size was in-fact sufficient (e.g., calculate $S_c$ and $\Delta$ from the developed model)
	
\item If estimand and criterion values fall short, evaluate the risk of bias (e.g., what is the biased posed to the prediction model if $\Delta>0.05$ or if $S_c<0.9$) 
\end{enumerate}

\subsubsection{Elaboration on select criteria}

\noindent Criterion (1): Whether the model in question is a calibration model or validation model may affect the parameter values. For example, in this case, $R=R_{adj}^{2}$.

\noindent Criterion (2-3): The principal objective of the prediction model is to provide an estimate for a particular estimand (e.g., the RMSE of a specific analyte). Thus, it is necessary that this objective be clearly articulated in the methods section, both as a matter of transparency and clarity, as well as a necessary component for the sample size criterion. Once the estimand has been identified, the necessary sample size criterion (and expression) should be presented. 

\noindent Criterion (5): Provide a short explanation as to the applicability of past literature on the proposed model development (i.e., how are they similar and how are they different). This will allow the reader to gauge whether they believe the basis for relevant parameters to be sufficient, and in the case that, post estimation, the sample size was insufficient, it will provide the chemometrician a candidate reason as to why the threshold value of the estimand was not achieved.

\noindent Criterion (7): From (i-vi), provide the sample size calculation accompanied by any additional assumptions that were made. Note that in some cases, a study may seek to estimate more than one estimand. In such cases, the sample size for each should be provided, and the largest of the two should be used. 

\noindent Criterion (8-9): By ascertaining the fitted values of the estimands and sample size criteria, the reader and chemometrician will be able to identify whether biases arising from optimism and overfitting were sufficiently overcome, and if not, to what extent they may still pose a problem. 

\subsubsection{Overview}
We reiterate that these are minimal guiding principles. However, they form the starting point for how sample size should be considered, calculated, and reported on. It is, in short, a framework by which transparent, rigorous models can be developed and applied, while providing assurance to the reader that steps to reduce bias were taken and allow the reader to make a judgement around the methodological limitations of a given study.

\subsection{Implications for chemometrics, planetary exploration and astrobiology }
While our article is applicable to chemometrics as a whole, there are specific challenges, which although not specific to space science and planetary exploration, are certainly prevalent in these fields. Specifically, we would like to draw attention to the case in which the calculated sample size is not feasible, the case in which complex models are used, and additional perspectives around sample size considerations for retrospectively curated datasets (i.e., where the data has already been collected). 

\subsubsection{Sample size feasibility}
In planetary exploration, a common type of sample is a geological standard, that is, a geological sample that has been rigorously characterized. While the range of costs for such standards may vary, price tags of \$100- \$300 are commonplace. Therefore, taking the empirical example provided in the previous section, depending on the criterion used, the cost for model calibration could have been as little as \$12,400 and as much as \$167,100. 

This is quite a large range, and the more likely cost would likely be somewhere in between; however, even a budget of say \$50,000 may be prohibitive for many endeavours. Should such an occassion arise, we reccomend that instead of relaxing the criteria (e.g., accepting a $\Delta$ of 0.1 or a $S_{c}$ of 0.85), the number of parameters should be reduced. This would ensure that the rigour of the model is maintained (e.g., the risk of overfitting is minimized) while allowing for the minimum sample size to be reduced, thereby reducing the costs. For example, in the previous section we assumed that $p=25$. If however, we were to work with a reduced set of parameters such that $p=15$, $n_{min}$ would be 151 based on the $\Delta$ criterion (compared to 251) and 140 based on the $S_{c}$ criterion (compared to 269). Reducing costs in this way can ensure that innovation in space science, both in academia and industry, may flourish, while maintaining the necessary level of rigour.

\subsubsection{Complex models}
The present work has focused on multivariable linear models, with some parenthetical extensions to more complex models. However, in chemometrics, for better or worse, models can be exceedingly complex, including those in planetary exploration (e.g., for Raman spectroscopy and LIBS) \cite{Clegg2017,LMM,veneranda2021raman,manrique2020evaluation,lopez2021raman} and in such cases, closed form solutions to the minimum sample size become untenable. Nevertheless, as we have discussed at length, arriving at the sample size is essential. We therefore propose that in situations where a closed form solutions is not possible or (as in the case of the shrinkage factor), sample size should be obtained by simulation. Specifically, we mean stochastic simulations, where both the independent and dependent variables are generated from some distribution (often hierarchically), the results of which are iteratively used to fit a model and compute estimands; this is a common approach in the biostatistical literature \cite{landau2013sample,teare2014sample,wynants2015simulation,hsieh1998simple,mundfrom2005minimum,morris2019using,feiveson2002power}.

\subsubsection{Prospective vs. Restrospective data}
We noted earlier on that the bulk of the present methods pre-suppose that we are interested in designing a prospective study (i.e., a study in which the data has not been curated yet). This is in contrast to a retrospective study, where data has already been collected and the data \say{is what it is}. The fact that the sample size calculations above are for a prospective study, give rise to a question, namely, should data from an already curated dataset be analyzed? We provide two answers here. Firstly, assuming that they data has not yet been analyzed, a sample size calculation can still be conducted, which would allow the chemometrician to gauge whether the dataset in question is likely to be at a risk of overfitting given some proposed model. Moreover, in so doing, the chemometrician will be able to determine what the maximum number of predictor parameters that can be included in a model is while minimizing the risk for overfitting. We emphasize here that this should only be done if the data has not been analyzed yet, otherwise, any number of biases may be introduced (beyond that of overfitting). 

It still begs the question, what if the number of candidate parameters has been reduced to what is believed to be the minimum acceptable set (e.g., by domain experts), can data still be analyzed even if the minimum sample size for the given $p$ is not satisfied? To which we and other authors \cite{hernan2021causal}, respond yes. The chemometrician can identify which, if any, estimands and thresholds are satisfied with the given sample size (e.g., the sample size may be sufficient on the basis of the $\Delta$ criterion, but not on the $S_{c}$ criterion.) It would then remain to report this transparently in the disseminated article, for example following the 9 steps presented above, and presenting a discussion into the likely limitations of the study, stating something along the lines of \say{this study is likely lacks a sufficient sample size to reliably mitigate the risk of over-fitting based on the $\Delta$ criterion}. On the whole, as long as the limitations induced by insufficient sample size are adequately addressed, there is no reason why a dataset would not be analyzed.

\section{Conclusion}
Sample size calculation is a pillar of evidence-based science. However, despite the widespread adoption of minimum sample size requirements in other fields, such requirements remain overwhelmingly absent in chemometrics. Through restructuring our discussion as a comprehensive communication piece, we have elucidated how certain statistical considerations in biomedical research criteria can serve a substantial role in the chemometrics landscape. This study marks a pivotal step in broadening interdisciplinary communication, underscoring the untapped potential in the convergence of medicine and chemometrics. The present work focuses on providing a theoretical basis. It worked examples of three criteria for minimum sample size calculations in linear regression with continuous outcomes – the $\Delta$ criterion, $S_{C}$ (or shrinkage) criterion, and the residual criterion (in addition to two extensions). We have provided a systematic framework for how sample size should be considered in developing a calibration model, and how it should be reported on. In short, this work is the first step in ensuring a clear, transparent, and evidence-based framework for the development and application of multivariable models in chemometrics. 

As we expand the search for life to different environments, the need for advanced chemometric models becomes more critical. The development of predictive models must therefore consider not just the immediate analytical challenges, but also the broader implications for astrobiology. Ensuring that our models are robust and our predictions accurate paves the way for one of humanity's most profound quests – to discover life elsewhere in the universe.

\section{Acknowledgments}
The authors would like to
acknowledge the financial support provided by the Natural Sciences and Engineering Research Council of Canada (NSERC) and the Canadian Space Agency (CSA). In addition, the authors would like to thank the Ministerio de Ciencia e Innovación of Spain for the funding of this project through grant PID2019-107442RB-C31.

\clearpage
\appendix
\input{Appendix_Sample_Size_includable}

\bibliographystyle{apalike}
\bibliography{Bib}
\end{document}

%% file: Appendix_Sample_Size_includable.tex

\section*{Abstract}
The purpose of this appendix is to provide clear definitions and explanations of key statistical concepts and parameters introduced in this research approach for astrobiology. The definitions outlined below are intended to aid readers, particularly those less familiar with statistical methodologies, in understanding the fundamental principles underlying the calculations and criteria presented in this study. 

We recognize that terms such as the adjusted \(R^2\), shrinkage factor, residual standard deviation, and root mean squared error (RMSE) may not be immediately accessible to all readers. Therefore, this appendix serves as a reference to improve the accessibility and clarity of the statistical methodology described in Sections 2 and 3 of the article. Each term is defined, with accompanying formulas and relevant references, to provide context and explain its role in determining the minimum sample size required for reliable regression results. This effort aims to ensure that the methodology is transparent and reproducible, aligning with the broader goals of this research.

\section*{1. Adjusted \(R^2\) (\(R^2_{\text{adj}}\))}
\textbf{Definition}: Adjusted \(R^2\) modifies the coefficient of determination (\(R^2\)) to account for the number of predictors included in the model. Unlike \(R^2\), which may increase simply by adding predictors, \(R^2_{\text{adj}}\) penalizes the addition of unnecessary predictors, making it a more reliable measure of model performance. \\
\textbf{Formula}:
\[
R^2_{\text{adj}} = 1 - \frac{(1 - R^2)(n - 1)}{n - p - 1}
\]
where:
\begin{itemize}
    \item \(n\): Total number of observations (sample size),
    \item \(p\): Number of predictor parameters.
\end{itemize}
\textbf{Significance}: \(R^2_{\text{adj}}\) is used to assess the risk of overfitting by calculating the difference (\(\Delta\)) between \(R^2\) and \(R^2_{\text{adj}}\):
\[
\Delta = R^2 - R^2_{\text{adj}}
\]
Small \(\Delta\) values (e.g., \(\Delta < 0.05\)) indicate minimal overfitting and an adequate sample size. \\
\textbf{References}: Steyerberg (2009), Riley et al. (2019), Copas (1983), Goldberger (1964).

\section*{2. Shrinkage Factor (\(S\))}
\textbf{Definition}: The shrinkage factor adjusts regression coefficients to account for overfitting. It scales down coefficients to make predictions more reliable for new data. The global shrinkage factor is denoted as \(S_c\). \\
\textbf{Formula}:
\[
S_c = 1 - \frac{p - 2}{LR}
\]
where:
\begin{itemize}
    \item \(LR = -2(l_{\text{null}} - l_{\text{model}})\): Likelihood ratio,
    \item \(l_{\text{null}}\): Log-likelihood of the intercept-only model,
    \item \(l_{\text{model}}\): Log-likelihood of the model with predictors.
\end{itemize}
\textbf{Threshold}: A shrinkage factor of \(S_c \geq 0.9\) is typically recommended to minimize overfitting. \\
\textbf{References}: Riley et al. (2019), Harrell Jr. (2015), Copas (1983, 1997).

\section*{3. Predictor Parameters and Variables}
\textbf{Definition}: Predictor parameters correspond to the coefficients estimated for each predictor in a model. It is critical to distinguish between:
\begin{itemize}
    \item \textbf{Variables}: Input features in the model (e.g., temperature, time),
    \item \textbf{Predictor Parameters}: Coefficients associated with each variable. For instance, a categorical variable with three levels (e.g., low, medium, high) would have two predictor parameters.
\end{itemize}
\textbf{Importance}: Accurately identifying \(p\) (the number of predictor parameters) is essential for sample size calculations. \\
\textbf{References}: Riley et al. (2019), van Smeden et al. (2019).

\section*{4. Residual Standard Deviation (\(\sigma\))}
\textbf{Definition}: Residual standard deviation quantifies the model's prediction error. It is the standard deviation of the residuals (differences between observed and predicted values). \\
\textbf{Formula}:
\[
\sigma = \sqrt{\frac{\sum_{i=1}^n (y_i - \hat{y}_i)^2}{n - p - 1}}
\]
where:
\begin{itemize}
    \item \(y_i\): Observed value for the \(i\)-th observation,
    \item \(\hat{y}_i\): Predicted value for the \(i\)-th observation.
\end{itemize}
\textbf{Applications}: Used in calculating confidence intervals and assessing the accuracy of regression coefficients. \\
\textbf{References}: Riley et al. (2019), Montgomery et al. (2021), Harrell Jr. (2015).

\section*{5. Root Mean Squared Error (RMSE)}
\textbf{Definition}: RMSE measures the accuracy of model predictions. It is the square root of the mean of the squared residuals. \\
\textbf{Formula}:
\[
RMSE = \sqrt{\frac{\sum_{i=1}^n (y_i - \hat{y}_i)^2}{n}}
\]
\textbf{Importance}: RMSE provides a direct assessment of how well the model predicts actual outcomes. Smaller RMSE values indicate better model performance. \\
\textbf{References}: Riley et al. (2019), Montgomery et al. (2021).

\section*{6. Optimism}
\textbf{Definition}: Optimism is the difference between a model’s performance in the training set (calibration set) and its expected performance on new data. High optimism often signals overfitting. \\
\textbf{References}: Steyerberg (2009), Riley et al. (2019).

\section*{7. Overfitting}
\textbf{Definition}: Overfitting occurs when a model learns noise in the training data as if it were a true signal, resulting in reduced predictive accuracy on unseen data. \\
\textbf{Indicators}:
\begin{itemize}
    \item Large \(\Delta\) values between \(R^2\) and \(R^2_{\text{adj}}\),
    \item Shrinkage factor (\(S\)) below 0.9.
\end{itemize}
\textbf{References}: Harrell Jr. (2015), Riley et al. (2019), Steyerberg et al. (2000).

\section*{8. Sample Size Criteria}
\textbf{Definition}: Criteria to determine the minimum number of observations required to ensure reliable model estimates. Key criteria include:
\begin{itemize}
    \item \(R^2\) vs. \(R^2_{\text{adj}}\): Absolute difference \(\Delta\) should be small (e.g., \(\Delta < 0.05\)),
    \item Shrinkage Factor (\(S\)): Should meet or exceed 0.9,
    \item Precision of the Residual Standard Deviation: Standard deviation estimated within a margin of error under a specific confidence interval.
\end{itemize}
\textbf{References}: Riley et al. (2019), van Smeden et al. (2019).